\documentclass[prb,aps,onecolumn,amsmath,amssymb,floatfix,
superscriptaddress]{revtex4}

\usepackage[dvips]{graphics}
\usepackage{color}
\definecolor{dred}{rgb}{0,0,0.6}

\begin{document}

\title{Modulation of circular current and associated magnetic field in 
a molecular junction: A new approach}

\author{Moumita Patra}

\affiliation{Physics and Applied Mathematics Unit, Indian Statistical
Institute, 203 Barrackpore Trunk Road, Kolkata-700 108, India}

\author{Santanu K. Maiti$^{*,}$}

\affiliation{Physics and Applied Mathematics Unit, Indian Statistical
Institute, 203 Barrackpore Trunk Road, Kolkata-700 108, India}

\begin{abstract}

A new proposal is given to control local magnetic field in a molecular 
junction. In presence of finite bias a net circular current is established
in the molecular ring which induces a magnetic field at its centre. 
Allowing a direct coupling between two electrodes, due to their close 
proximity, and changing its strength we can regulate circular current as 
well as magnetic field for a wide range, without disturbing any other 
physical parameters. We strongly believe that our proposal is quite robust 
compared to existing approaches of controlling local magnetic field and 
can be verified experimentally.

\end{abstract}

\maketitle

The study of electronic transport through single molecules has been the
objects of intense research due to the fact that molecular components can
be utilized as significant functional elements in electronic devices.
In 1974 Aviram and Ratner~\cite{ref1} first proposed a unimolecular device 
considering a molecule as the basic building block, and latter many works 
have been done~\cite{ref2,ref3,ref4,ref5,ref6,ref7,ref8,ref9,ref10,
ref11,ref12,ref13,ref14,ref15,ref16,ref17} to explore electron 
transport through different simple as well as complex molecular structures.

Though a wealth of literature knowledge has been established in the field
of molecular transport, most of the works have focused essentially on net 
junction current, while very few attempts have been made~\cite{cir1,cir2,
cir3,cir4,cir5,cir6,cir7,cir8,cir9} so far where distribution of current 
in different arms of a molecular junction has been analyzed. 
\begin{figure}[ht]
{\centering \resizebox*{6cm}{5.5cm}{\includegraphics{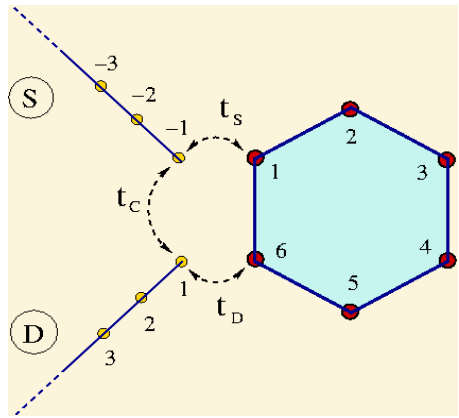}}\par}
\caption{(Color online). Schematic view of a molecular junction where a 
benzene molecule is coupled to source (S) and drain (D) electrodes in 
ortho-configuration. Due to close proximity there exists a {\em direct 
coupling} between S and D which provides a {\em new path} along with the
conventional path i.e., the bridging molecule.}
\label{f1}
\end{figure}
In presence
of finite bias a net circular current is established in the molecular ring 
which results a non-zero magnetic field at its centre. Depending on bias
voltage and molecule-to-electrode interface geometry this magnetic field 
becomes quite high and in some cases it becomes $\sim\,$Millitesla (mT) or 
even $\sim\,$T~\cite{cir7,cir8}. A number of recent investigations of 
electronic transport through molecular 
junctions~\cite{cir7,cir8,nrf1,nrf2,nrf3} have 
shown that in the limit of weak molecule-to-electrode coupling much higher 
circular current is obtained in molecular loops compared to the net transport 
current across the junction. Possible applications of such high {\em local} 
magnetic fields in molecular systems came into limelight following the 
realization of controlling spin orientation~\cite{cir3} of a cation site 
embedded in a conducting junction by the local magnetic field induced by
loop current or the prediction 
of carbon nanotubes as molecular solenoids~\cite{nrf3,ctube1,ctube2}.
Considering a T-shape tape-porphyrin molecular wire Tagami and Tsukada have 
shown that the current which is established in the molecular loop produces
the local magnetic field $\sim 0.1\,$T at the bias voltage of $1.2\,$V, that 
can be utilized to regulate local spin orientation, and it has an important
viewpoint as detecting the spin orientation by means of changing the bias 
polarity one can get a clear idea of the existence of circular current in 
the molecular loop. The phenomenon of circular current is also directly 
linked with other context the so-called current transfer process~\cite{ct}, 
where a current imposed in one path affects a current in other arms exploring
the quantum interference affect, which certainly demands a detailed 
analysis.

Though several suggestions were made for the possible exploitations of
such high local magnetic fields at the molecular regime, probably the 
most significant application can be the generation of spin-based quantum
computers~\cite{qc1,qc2,qc3,qc4}. To achieve this goal proper spin 
regulation is highly important, which on the other hand requires finite 
tuning of magnetic field in a localized region. Few propositions have
already been done along this direction. For instance, using phase locked 
infra-red laser pulses Pershin and Piermarocchi have shown~\cite{Per} that 
circular current can be established in an isolated quantum ring
where the magnetic field reaches up to {\em few mT}. Utilizing this local
magnetic field they have shown how the spin orientation, provided e.g.,
by a magnetic impurity embedded at the ring centre or on top of a ring,
can be locally controlled by magnetic field due to the current in the
ring. In other work Lidar and Thywissen have established~\cite{lidar} 
that a localized magnetic field, which may reach up to $10\,$mT, can be 
generated with the help of {\em an infinite array of parallel current 
carrying wires}, though it has severe limitation due to heating effect 
and one has to work at much lower temperature ($< 2.4\,$mK). Comparing 
all these propositions we can argue that bias induced magnetic field, 
associated with circular current, in a nano-junction is quite robust and 
easy to operate~\cite{cir7,cir8,mag1,mag2,mag3}. The essential motivations
behind the consideration of a molecular junction with loop structure(s) are 
as follows: (a) Bias induced circular current produces strong magnetic 
field (that can also be varied in a wide range) at the molecular/nano-scale 
level compared to the net junction current. At this length scale simple 
quantum wire cannot produce such a strong magnetic field. (b) Exploiting 
quantum interference effect several anomalous features can be observed in 
ring-like geometry, which are not possible in conducting junctions without 
any loop. (c) Spectral response of magnetic ions placed near or on the 
molecular ring to the current induced magnetic field gives an atypical 
observation of magnetic shielding and deshielding effect in NMR spectra 
of aromatic molecules~\cite{nmr}. (d) Another operation can also be 
implemented by assigning up and down spin states as two binary logic bits
$0$ and $1$. The flipping of spin states, as a result of 
local magnetic field will correspond to the switching between $0$ and $1$
states, which thus carry quantum information. This is the basic principle
used in designing quantum computation which reduces much power dissipation,
compared to the conventional computing which is charge based where two 
different charges are assigned to encode binary logic bits $0$ and $1$, and
involves charge flow that costs excessive power loss. Thus, the study of 
circular current due to voltage bias in the molecular scale level is 
certainly worthy and interesting.

In the present paper we essentially focus on how to control circular
current and associated magnetic field in a molecular junction having single
or multiple loops coupled to source and drain electrodes. Due to close
proximity electrons can directly hop between the end atomic sites of these
two electrodes and tuning this coupling strength, which is done simply by
changing the orientation/position of these electrodes, {\em we can regulate 
circular current and associated magnetic field in a wide range}. 
{\em No one has addressed this issue before, to the best of our concern, 
and certainly gives a new insight to modulate electron transmission through 
a nano-junction.}

\vskip 0.5cm
\noindent
{\bf Molecular Model and Theoretical Framework}
\vskip 0.15cm
\noindent
The molecular junction is schematically shown in Fig.~\ref{f1} where a 
benzene molecule is coupled to two one-dimensional (1D) perfect electrodes,
viz, source and drain. The electrodes are connected to the molecule in 
ortho-configuration such that an electron can directly hop between the end
atomic sites of these electrodes due to their close proximity which 
essentially provides a {\em new path} in addition to the conventional path 
i.e., the molecular ring.

To describe this model we use tight-binding (TB) framework, which is most
convenient for analyzing electron transport through a molecular junction
particularly in the limit of non-interacting electrons. The full Hamiltonian
of the molecular junction can be written as: $H=H_M+H_S+H_D+H_T$, where 
$H_M$ corresponds to the Hamiltonian for the molecule, $H_{S(D)}$ represents
the Hamiltonian for the source(drain) electrode and $H_T$ gives the tunneling
Hamiltonian. In TB framework these sub-Hamiltonians are expressed as follows:
\begin{equation}
H_M=\sum\limits_{i=1} \epsilon_i c_i^{\dagger} c_i + \sum\limits_{i=1} t
\left(c_{i+1}^{\dagger} c_i + c_i^{\dagger} c_{i+1} \right)
\label{eq1}
\end{equation}

\begin{equation}
H_S=\sum\limits_{n \le -1} \epsilon_0 a_n^{\dagger} a_n + 
\sum\limits_{n \le -1} t_0 \left(a_n^{\dagger} a_{n-1} + 
a_{n-1}^{\dagger} a_n \right)
\label{eq2}
\end{equation}

\begin{equation}
H_D=\sum\limits_{n \ge 1} \epsilon_0 b_n^{\dagger} b_n + 
\sum\limits_{n \ge 1} t_0 \left(b_n^{\dagger} b_{n+1} + 
b_{n+1}^{\dagger} b_n \right)
\label{eq3}
\end{equation}

\begin{eqnarray}
H_T &=& t_S \left(c_p^{\dagger}a_{-1} + a_{-1}^{\dagger} c_p \right)
+ t_D \left(c_q^{\dagger}b_1 + b_1^{\dagger} c_q \right) \nonumber \\
& + & t_c \left(a_{-1}^{\dagger} b_1 + b_1^{\dagger} a_{-1} \right)
\label{eq4}
\end{eqnarray}
Here $c_i^{\dagger}$ and $c_i$ correspond to the creation and annihilation
operators, respectively, for an electron at $i$-th site of the molecular ring,
while these operators are $a_n^{\dagger}$, $a_n$ and $b_n^{\dagger}$, $b_n$
for the source and drain electrodes, respectively. The molecule is 
characterized by the on-site potential $\epsilon_i$ and nearest-neighbor 
hopping integral $t$, whereas for the side attached electrodes these 
parameters are $\epsilon_0$ and $t_0$, respectively. $t_S$ describes the
molecular coupling with the source and it is $t_D$ for the drain. These
electrodes are connected at the sites $p$ and $q$ (which are variable and
nearest-neighbors). For this molecular junction (Fig.~\ref{f1}) $p=1$ and 
$q=6$. $t_c$ represents the inter-electrode coupling and it can be tuned 
either by changing the separation between the electrodes or by rotating them. 
Our main concern in this article is how $t_c$ affects electronic transmission
through the molecular junction.

To evaluate transmission probability across the molecular wire we adopt
wave-guide theory~\cite{cir7,cir8,wg1,wg2,wg3} where a set of coupled linear 
equations involving wave amplitudes at different lattice sites are solved. 
These coupled equations are generated from the Schr\"{o}dinger equation 
$H|\psi\rangle = E|\psi\rangle$, considering $|\psi\rangle$ in the form:
\begin{equation}
|\psi\rangle =\left[\sum\limits_{n \le -1}A_n a_n^{\dagger} +
\sum\limits_{n \ge 1}B_n b_n^{\dagger} +
\sum\limits_{i=1}C_i c_i^{\dagger}\right]|0\rangle
\label{eq5}
\end{equation}
where $A_n$, $B_n$ and $C_i$ correspond to the amplitudes for an electron
at site $n$ of the source/drain electrode and at the site $i$ of the ring,
respectively. In terms of the reflection and transmission coefficients $r$
and $\tau$, the amplitudes $A_n$ and $B_n$ can be written as 
$A_n=e^{ik(n+1)} + r e^{-ik(n+1)}$ and $B_n=\tau e^{ikn}$, where we assume
that a plane wave with unit amplitude is coming from the source. Thus for
each wave vector $k$, associated with energy $E$, we calculate $\tau$
from the set of linear equations and get the transmission probability 
\begin{equation}
T(E)=|\tau|^2=|B_1|^2.
\label{eq6}
\end{equation}
Using the transmission function $T(E)$, net junction current at absolute 
zero temperature for a particular voltage bias $V$ is determined from the 
relation~\cite{datta}
\begin{equation}
I_T(V) = \frac{e}{\pi \hbar} \int\limits_{E_F-\frac{eV}{2}}^{E_F+\frac{eV}{2}}
T(E) \, dE
\label{eq7}
\end{equation}
where $E_F$ is the equilibrium Fermi energy.

Now to find circular current in the molecular ring we need to calculate 
current carried by individual bonds. For any such bond, connecting the
sites $i$ and $i+1$, it becomes~\cite{cir7,cir8}
\begin{equation}
I_{i,i+1}(V) = \int\limits_{E_F-\frac{eV}{2}}^{E_F+\frac{eV}{2}}
J_{i,i+1}(E) \, dE
\label{eq8}
\end{equation}
where $J_{i,i+1}=(2e/\hbar) \mbox{Im}\left[t\,C_i^*C_{i+1} \right]$.
With these bond currents the net circular current is calculated from the
relation~\cite{cir7,cir8,mag1,mag2,mag3}
\begin{equation}
I_c=\frac{1}{L} \sum_i I_{i,i+1}\, a
\label{eq9}
\end{equation}
where $L=Na$, $a$ being the lattice spacing and $N$ represents the total 
number of atomic sites in the bridging molecule. We assign a positive sign
to a current flowing in the anti-clockwise direction.

Due to this circular current a net magnetic field is established. The local
magnetic field at any arbitrary point $r$ (say) inside the molecule can be 
determined using the Biot-Savart's law~\cite{cir7,cir8,mag1,mag2,mag3}
\begin{equation}
\vec{B}(\vec{r}) = \sum\limits_{\langle i,j \rangle} \left(\frac{\mu_0}{4\pi}
\right)
\int I_{i,j}\frac{d\vec{r^{\prime}} \times(\vec{r}-\vec{r^{\prime}})}
{|\vec{r}-\vec{r^{\prime}})|^3}
\label{bb}
\end{equation}
where $\mu_0$ is the magnetic constant.

\vskip 0.5cm
\noindent
{\bf Results and Discussion}
\vskip 0.15cm
\noindent
Based on the above theoretical framework now we present our results which
include two-terminal transmission probability, junction current, circular
current and associated magnetic field at the ring centre. There are some
physical parameters those values are kept constant throughout the numerical
calculations. These parameter are described as follows. In the 
molecular ring we choose $\epsilon_i=-1.5\,$eV and $t=2.5\,$eV, while for 
the side-attached electrodes they are: $\epsilon_0=0$ and $t_0=3\,$eV. 
The molecule-to-electrode coupling parameters ($t_S$ and $t_D$) are fixed 
at $1\,$eV, and the equilibrium Fermi energy $E_F$ is 
set at zero. The values of other physical parameters, those are not same 
for all figures, are specified in appropriate places. To calculate magnetic
field we assume the perpendicular distance from the centre of the benzene
\begin{figure}[ht]
{\centering \resizebox*{6cm}{4cm}{\includegraphics{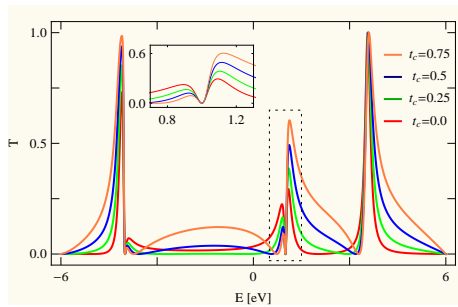}}\par}
\caption{(Color online). Two-terminal transmission probability ($T$) as 
a function of energy ($E$) for the molecular junction (shown in Fig.~\ref{f1})
at some typical values of $t_c$. The inset of the figure 
represents the zoomed version of the dashed framed region. A sharp dip
(vanishing transmission) is observed at $\sim E=1\,$eV and across this dip
resonant peaks exhibit completely opposite scenario. In the right side the
peak height gradually increases with $t_c$, while it decreases with $t_c$ in
the other side of the dip. The similar feature is also observed in other
energies where a transmission peak is followed by a dip.}
\label{f2}
\end{figure}
ring to any $C$-$C$ bond is $\sim0.13\,$nm~\cite{cir8}. 

Before going to the central part of our analysis i.e., how {\em direct 
coupling} ($t_c$) affects circular current and induced magnetic field, 
\begin{figure}[ht]
{\centering \resizebox*{6cm}{4cm}{\includegraphics{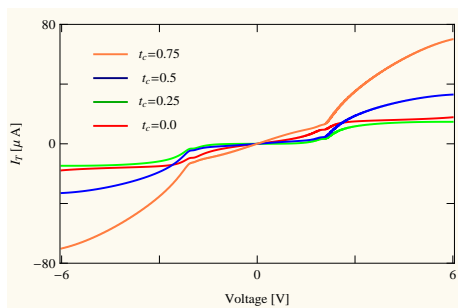}}\par}
\caption{(Color online). Junction current $I_T$ as a function of applied 
bias voltage $V$ for the ortho-connected benzene molecule considering 
identical values of $t_c$ as taken in Fig.~\ref{f2}.}
\label{f3}
\end{figure}
\begin{figure}[ht]
{\centering \resizebox*{6cm}{4cm}{\includegraphics{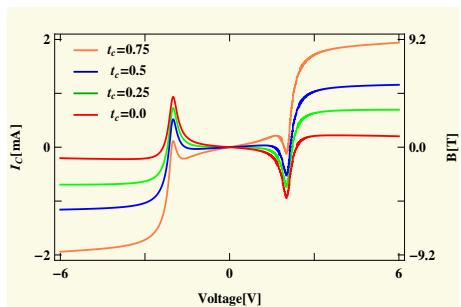}}\par}
\caption{(Color online). Circular current $I_c$ and associated magnetic field
$B$ at the centre of the benzene molecule as a function of bias voltage 
for different values of $t_c$.}
\label{f4}
\end{figure}
let us focus on transmission probability and junction current. In 
Fig.~\ref{f2} we present the variation of two-terminal transmission 
probability as a function of injecting electron energy for the benzene 
molecule considering some typical values of $t_c$. For $t_c=0$ fine 
resonant peaks associated with energy eigenvalues of the molecular ring are 
obtained while the transmission probability drops very close to zero for 
all other energies. This behavior has already been discussed in several 
earlier papers~\cite{ref12,ref17,cir7} for ortho-connected benzene ring. 
The situation becomes very interesting when we include the 
effect of $t_c$. Apparently it shows that electron gets transmitted almost 
for the entire energy window and the transmission amplitude gradually 
increases with the rise of $t_c$. But, a careful inspection yields many
fascinating points. To reveal this fact we select a small part of the 
spectrum, the dashed framed region, and place its zoomed version in the 
inset. Very 
interestingly we see that a sharp dip (vanishing transmission) appears
at $\sim E=1\,$eV, and above and below this dip resonant curves exhibit 
\begin{figure}[ht]
{\centering \resizebox*{6.5cm}{4cm}{\includegraphics{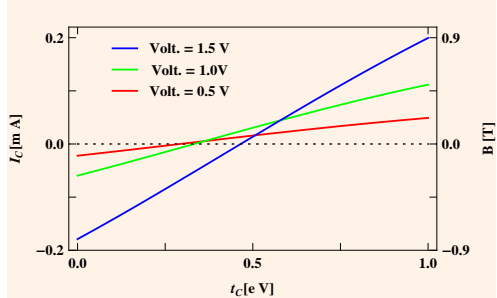}}\par}
\caption{(Color online). Dependence of $I_c$ and $B$ with $t_c$ for the
benzene molecule at some typical bias voltages. The dashed horizontal line
represents the line of zero circular current.}
\label{f5}
\end{figure}
completely opposite behavior. One side of this dip, the height of the 
peak increases while in the other side it gradually decreases with respect 
to the coupling parameter $t_c$. This feature is also observed in other 
energies where a transmission peak is followed by a dip. It is an 
important observation since one can get higher and/or lower electronic
transmission at different energies simply by tuning the external coupling
parameter $t_c$, without changing any other physical variables. The 
anomalous feature in this ring-like geometry is observed due to the presence
of the {\em new path} between the electrodes. A combined interference effect
among electronic waves passing through different arms (upper and lower arms
of the molecular ring including the external new path) leads to such a nice 
\begin{figure}[ht]
{\centering \resizebox*{7cm}{7.5cm}{\includegraphics{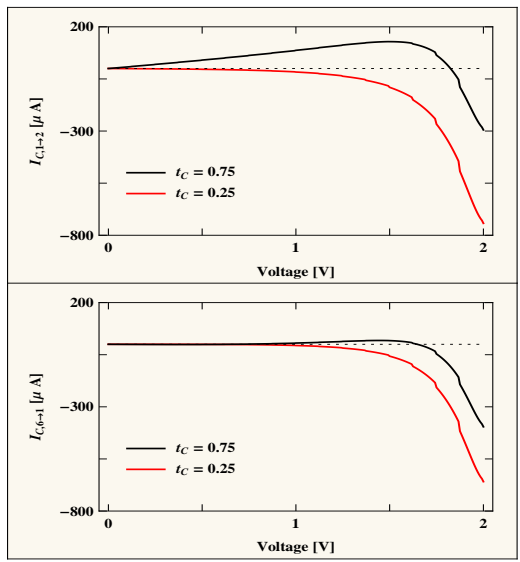}}\par}
\caption{(Color online). Bond currents (two bonds are taken
from two arms of the molecular junction) as a function of bias
voltage for the benzene molecule at two different values of $t_c$. We choose
these two typical values of $t_c$ to explore the sign reversal of circular
current and induced magnetic field displayed in Fig.~\ref{f5}. The dashed
horizontal line represents the line of zero bond current.}
\label{bondv}
\end{figure}
phenomenon, and of course would not be noticed in molecular junctions 
without any loop structure. Thus a competition takes place between the 
interfering paths i.e., the molecular arms and the external path, and the
response depends on the resultant of all these paths. For strong enough
$t_c$ electrons mostly follow the external path, avoiding the conventional
molecular ring.

The above signature is clearly reflected in the current-voltage 
characteristics as 
the junction current is evaluated by integrating transmission function $T$ 
over an energy window associated with the bias voltage $V$ (Eq.~\ref{eq7}).
Figure~\ref{f3} displays the dependence of junction current $I_T$ with 
applied bias voltage $V$ for the othro-connected benzene molecule for some
specific values of $t_c$. The current starts increasing approximately linearly
with $V$, while a sudden change of its amplitude takes place at a critical 
voltage ($V\sim2\,$V). This sudden jump is associated with the crossing
\begin{figure}[ht]
{\centering \resizebox*{5cm}{5cm}{\includegraphics{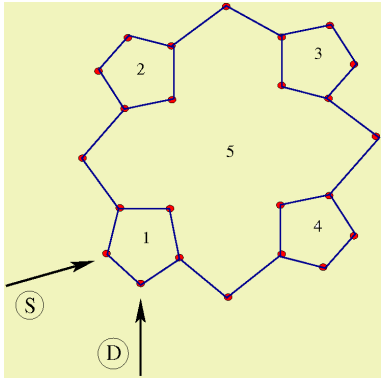}}\par}
\caption{(Color online). Schematic view of another molecular
junction where the benzene molecule is replaced by porphyrin molecule. The
ring-to-electrode configuration remains same as in Fig.~\ref{f1}. The 
numbers $1$, $2 \ldots$, $5$ represents the loop numbers.}
\label{por}
\end{figure}
of one of the resonant energy levels which is clearly seen from the $T$-$E$ 
spectrum (Fig.~\ref{f2}). Most interestingly we see that for low enough 
$t_c$ current is smaller (green line) compared to the molecular junction
without any $t_c$ (red line), but eventually the current increases sharply 
with $t_c$, following the $T$-$E$ curve (Fig.~\ref{f2}). 

Now we concentrate on the variations of circular current and induced magnetic 
\begin{figure}[ht]
{\centering \resizebox*{9cm}{11cm}{\includegraphics{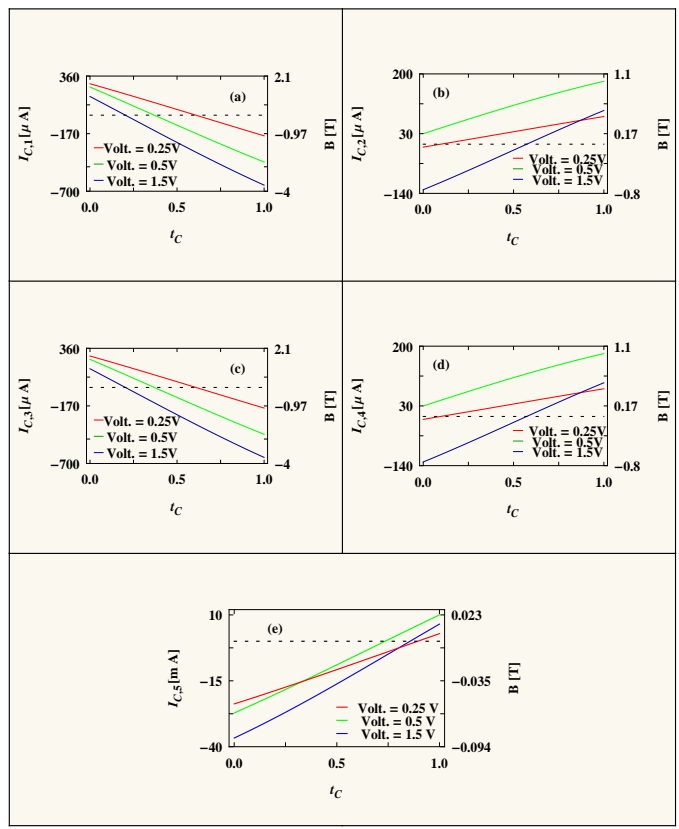}}\par}
\caption{(Color online). Circular current ($I_{c,n}$, $n=1$, 
$2\ldots$, $5$) and induced magnetic field ($B$) in different sub-loops 
(presented in (a)-(e)) of the molecular junction shown in Fig.~\ref{por} 
as a function of $t_c$ for some typical bias voltages.}
\label{fn}
\end{figure}
field produced at the ring centre for the molecular junction given in 
Fig.~\ref{f1}. The results are presented in Fig.~\ref{f4}. Unlike junction
current ($I_T$), circular current ($I_c$) changes its sign in different
voltage regimes for any side (positive or negative) of the applied bias. 
And also the magnitude of $I_c$ may be sufficiently large compared to the 
transport current $I_T$, depending on the external voltage $V$. For narrow
voltages when no resonant energy level appears 
within the voltage window we get vanishing circular current. While a non-zero 
contribution comes when anyone of such energy levels lies within the voltage 
window. With increasing voltage more and more resonant energy levels appear 
within the window and all of them contribute to the current, resulting a net 
circular current which may be positive or negative depending of the sign of 
the dominating energy levels (the sign reversal can be clearly understood 
from the forthcoming analysis). Though junction 
current always increases with voltage bias in conventional conducting 
junctions (where negative differential resistance effect is not considered). 
Most importantly, the magnetic field which is developed at the ring centre 
as a result of 
this circular current is surprisingly high, and it increases significantly 
with $t_c$. For a wide voltage region ($>\sim2\,$V) the magnetic field remains 
almost constant for any specific $t_c$ (Fig.~\ref{f4}), following $I_c$, as 
within this window there is no other energy channel to contribute current.

In order to see more clearly the dependence of circular current and 
associated magnetic field on $t_c$ in Fig.~\ref{f5} we present their 
variations as a function of $t_c$ for some typical values of bias voltage 
$V$. It is observed that both the circular current and induced magnetic 
field decrease with $t_c$ and reaching to zero, and eventually they increase 
with increasing the coupling parameter $t_c$. For lower $t_c$, one of the 
doubly degenerate energy levels comes within the voltage window (the 
degeneracy disappears as a result of molecule-to-electrode coupling) which 
contributes to the current. But as we increase $t_c$ the other resonant 
energy levels also appear within this bias window and contributes current 
in opposite direction with respect to 
the earlier one yielding a reduction of current. Finally, when they 
become exactly opposite with each other a vanishing net current is obtained. 
Beyond this critical value of $t_c$ both these states contribute in the 
same direction providing a resultant higher circular current. 
From this behavior it can be manifested that {\em tuning the coupling
between source and drain electrodes one can regulate circular current and
thus locally control induced magnetic field for a wide range starting from 
zero to few Tesla. Certainly this phenomenon gives a new way of controlling 
magnetic field in a specific region without disturbing any physical
parameters of the system and can be utilized in designing effective spin 
based quantum devices.}

The sign reversal of circular current, and hence the 
induced magnetic field, with $t_c$ for different voltages can be clearly 
understood from the variation of bond currents, as circular current is 
determined from the bond currents (Eq.~\ref{eq9}). The variations of two bond 
currents, where the bonds are chosen from the two arms of the junction, with
bias voltage are shown in Fig.~\ref{bondv}. The results are computed for
two typical values of $t_c$, one for which $I_c$ is negative in 
Fig.~\ref{f5} ($t_c=0.25\,$eV) and for the other ($t_c=0.75\,$eV) 
$I_c$ becomes positive in Fig.~\ref{f5}. From the spectra given in 
Fig.~\ref{bondv} it is clearly noticed that for a fixed $t_c$ bond current 
changes its sign with voltage. At the same footing for a fixed bias voltage
the sign reversal of bond current also takes place with the change of $t_c$
yielding a change of sign of the circular current $I_c$.

The results presented above are worked out for the molecular wire containing
only the benzene molecule. So the question naturally comes whether similar
kind of behavior is observed in other molecular wires with higher number of
loops. To answer it now we analyze the behavior of circular current and 
associated magnetic field of other relevant molecular
structure, namely, porphyrin, that is connected with the source and drain 
electrodes as prescribed in Fig.~\ref{por}, analogous to the configuration 
given in Fig.~\ref{f1}. The results are shown in Fig.~\ref{fn}. Qualitatively 
the circular currents and induced magnetic fields in different sub-loops of
the porphyrin molecule exhibit almost similar characteristic features to
what we get for the case of benzene molecule (Fig.~\ref{f5}). An additional 
important feature is that in some wide voltage regions circular currents in 
different loops are opposite in sign. Note that the magnitude of circular 
current in the four outer loops is much larger compared to the bigger inner 
one. Here it is relevant to note that based on circular current induced 
magnetic field, controlling of spin orientation of a cation site embedded
in the T-shape tape-porphyrin molecular wires has already been discussed 
elaborately in Ref.~\cite{cir3}. Identical features are also obtained for 
other molecular junctions involving several such molecular loops (not shown 
here) which we confirm through our detailed numerical calculations.
{\em Thus, one can in principle consider a molecular system where magnetic 
fields of variable strengths can be established in different sub-regions 
of the geometry that might be very helpful for designing nanoelectronic 
quantum devices.}

\vskip 0.5cm
\noindent
{\bf Summary}
\vskip 0.15cm
\noindent
In this work we have demonstrated how to control local magnetic field in
a wide region (from zero to a surprisingly large value) considering a 
simple molecular structure by introducing a new path between two electrodes. 
Using the wave-guide theory, we have calculated two-terminal transmission
probability, junction current, circular current and current induced magnetic 
field at ring centre(s) based on a coherent tight-binding framework. Our 
finding, to the best of our concern, gives a unique idea of regulating 
electron transport through a conducting junction.

\vskip 0.5cm
\noindent
{\bf Acknowledgement}
\vskip 0.15cm
\noindent
MP is thankful to University Grants Commission (UGC), India 
(F. 2-10/2012(SA-I)) for research fellowship.

\vskip 0.5cm
\noindent
{\bf Author Contributions}
\vskip 0.15cm
\noindent
S.K.M. conceived the project. M.P. performed numerical calculations.
M.P. and S.K.M. analyzed the data. S.K.M. supervised the theoretical 
calculations. M.P. and S.K.M. co-wrote the paper.

\vskip 0.5cm
\noindent
{\bf Additional Information}
\vskip 0.15cm
\noindent
Correspondence should be addressed to S.K.M.

\vskip 0.5cm
\noindent
{\bf $^*$Correspondence to:}
santanu.maiti@isical.ac.in

\vskip 0.5cm
\noindent
{\bf Competing financial interests}
\vskip 0.15cm
\noindent
The authors declare no competing financial interests.

\end{document}